\documentclass[twocolumn,showpacs,prb]{revtex4}
\usepackage{bm}
\usepackage{epsfig}

\begin{document}
\title{Spin Splitting Induced Photogalvanic Effect in Quantum Wells}

\author{L.E.~Golub}

\email{golub@coherent.ioffe.rssi.ru}

\affiliation{A.F.~Ioffe Physico-Technical Institute, Russian
Academy of Sciences, 194021 St.~Petersburg, Russia}


\begin{abstract}
A theory of the circular photogalvanic effect caused by spin
splitting in quantum wells is developed. Direct interband
transitions between the hole and electron size-quantized subbands
are considered. The photocurrent excitation spectrum is shown to
depend strongly on the form of the spin-orbit interaction. In the
case of structure inversion asymmetry induced (Rashba)
spin-splitting, the current is a linear function of light
frequency near the absorption edge, and for the higher excitation
energy the spectrum changes its sign and has a minimum. In
contrast, when the bulk inversion asymmetry (Dresselhaus
splitting) dominates, the photocurrent edge behavior is parabolic,
and then the spectrum is sign-constant and has a maximum.
\end{abstract}

\maketitle

\section{Introduction}
Spin properties of carriers attract great attention in recent
years due to rapidly developed spintronics dealing with
manipulation of spin in electronic devices.~\cite{spintronics} The
first idea has been put forward by Datta and Das, who proposed a
spin field-effect transistor.~\cite{spin_transistor} Its work is
based on a change of the Rashba field in semiconductor
heterostructures, caused by structure inversion asymmetry (SIA).

The promising materials for spintronics are III-V semiconductor
quantum wells~(QWs) whose spin properties are well documented and
can be controlled by advanced technology. However, in addition to
the Rashba spin-orbit interaction, another effective magnetic
field acts on carriers in zinc-blende heterostructures. This is
the so-called Dresselhaus field caused by bulk inversion asymmetry
(BIA).~\cite{Silva92}

Both BIA and SIA give rise to many spin-dependent phenomena in
QWs, such as an existence of beats in the Shubnikov-de~Haas
oscillations,~\cite{SdHspin} spin relaxation,~\cite{DK} splitting
in polarized Raman scattering spectra,~\cite{Jusserand92} and
positive anomalous magnetoresistance.~\cite{WLspin} Spin
splittings and relaxation times have been extracted from these
experiments. However, in [001]~QWs, the BIA and SIA spin-orbit
interactions result in the same dependences of spin splittings and
spin relaxation times on the wave vector. Therefore it is
impossible to determine the nature for the spin splitting.
Only in the simultaneous presence of both BIA and SIA of
comparable strengths, one can observe new effects, see
Ref.~\onlinecite{Review} and references therein. However the
latter situation is rare in real systems because it requires a
special structure design.

In this work, the other spin-dependent phenomenon is investigated
which is essentially different from the mentioned above. This is
the circular photogalvanic effect which is a conversion of photon
angular momentum into a directed motion of charge carriers. This
leads to an appearance of electric current under absorption of a
circularly-polarized light.\cite{ELGE} The photocurrent reverses
its direction under inversion of the light helicity.
Microscopically, the circular photocurrent appears owing to a
coupling between orbital and spin degrees of freedom. In
semiconductors the coupling is a consequence of the spin-orbit
interaction. In two-dimensional systems, the circular
photogalvanic effect can be caused by both Rashba and Dresselhaus
effective magnetic fields. In this paper we show that SIA and BIA
result in experimentally distinguishable photocurrents.

Recently started activity on circular photogalvanics in QWs
attracted big attention.~\cite{Ganichev_last,Ganichev_Nature} The
photocurrents induced by both BIA and SIA have been investigated.
The circular photocurrent has been mostly studied under {\em
intra}band optical transitions induced by infrared or far-infrared
excitations. However it is important to extend the studies on the
optical range where the effect is expected to be much stronger. In
this case, the photocurrent appears due to {\em inter}band
transitions.

In the present work, the theory of  the interband circular
photogalvanic effect in QWs is developed, and the photocurrent
spectra are calculated.

\section{Theory}

Spin splittings of electron or hole subbands give rise to the
circular photogalvanic effect. In order to obtain non-zero
photocurrent, it is enough to include spin-orbit interaction for
only one kind of carriers. Here we take into account spin-orbit
splitting in the conduction band. In QWs, the spin-orbit
interaction is described by the linear in the wave vector
Hamiltonian
\begin{equation}
\label{H} H({\bm k}) = \beta_{il} \sigma_i k_l \:,
\end{equation}
where $\sigma_i$ are the Pauli matrices.

Tensor $\beta$ is determined by the symmetry of the QW. As
mentioned in Introduction, in structures with a zinc-blende
lattice, there is a contribution due to BIA known as the
Dresselhaus term. For [001]~QWs, it has two non-zero components,
namely
\begin{equation}
\label{BIA} \beta_{xx} = - \beta_{yy} \equiv \beta_{\rm BIA} \:,
\end{equation}
where $x || [100], y || [010]$, and the $z$-axis coincides with
the growth direction.

SIA appears due to inequivalence of the right and left interfaces
of the QW, electric fields applied along $z$ direction, etc. It
leads to an additional contribution to the spin-orbit Hamiltonian,
the so-called Rashba term:
\begin{equation}
\label{SIA} \beta_{xy} = - \beta_{yx} \equiv \beta_{\rm SIA} \:.
\end{equation}

In the presence of the spin-orbit interaction~(\ref{H}), the two
electron states with a given wave vector $\bm k$ have a splitting
$\Delta = 2 \sqrt{(\beta_{xl} k_l)^2 + (\beta_{yl} k_l)^2}$. We
denote these states by the index $m = \pm$. Their envelope
functions in size-quantized subbands can be chosen in the form
\begin{equation}
\label{el_w_f} |m\rangle = \exp{(i \: \bm{k} \cdot \bm{\rho})} \:
\varphi(z) \left[ {m \over \sqrt{2}} \exp{(- i \Phi_{\bm k})}
\uparrow + {1 \over \sqrt{2}} \downarrow  \right] \:,
\end{equation}
where the phase $\Phi_{\bm k}$ is given by
$$\tan{\Phi_{\bm k}} = {\beta_{xl} k_l \over \beta_{yi} k_i},$$
$\uparrow$ and $\downarrow$ are the spin-up and spin-down states,
and $\bm{\rho}$ is the electron position in the plane of QW. For
odd (even) subbands of size-quantization, $\varphi(-z) = \pm
\varphi(z)$, respectively.

The two hole states in the same subband and with the same ${\bm
k}$ are assumed to be degenerate. We denote their dispersion,
calculated in the spherical approximation, as $E_h(k)$. The states
can be chosen as symmetrical ($s$) and antisymmetrical ($a$) in
relation to the mirror reflection in the plane located in the
middle of the QW. The corresponding wave functions have the
form~\cite{MPP}
\begin{eqnarray}
\label{h_w_f} && |s\rangle = \exp{(i \: \bm{k} \cdot \bm{\rho}
\:)}\\
&& \times N \left\{ - \left[ u_{3/2} + \sqrt{3} W_+ \exp{(2 i
\phi_{\bm k})} u_{-1/2} \right] C(z) \right.  \nonumber
\\
&&+ \left.  i \zeta \left[ \exp{(3 i \phi_{\bm k})} u_{-3/2} +
\sqrt{3} W_- \exp{(i \phi_{\bm k})} u_{1/2} \right] S(z)\right\},
\nonumber
\end{eqnarray}
and $|a\rangle$ can be obtained from $|s\rangle$ by applying the
operations of time- and space-inversion. Here $u_{\pm 3/2}$,
$u_{\pm 1/2}$ are the Bloch functions at the top of the valence
band, $C$ and $S$ are, respectively, even and odd functions of
coordinate $z$,~\cite{zeta} and $\phi_{\bm k}$ is the angle
between [100] and ${\bm k}$:
$$\tan{\phi_{\bm k}} = k_y/k_x.$$

\begin{figure}
\epsfxsize=3in \epsfysize=3in \centering{\epsfbox{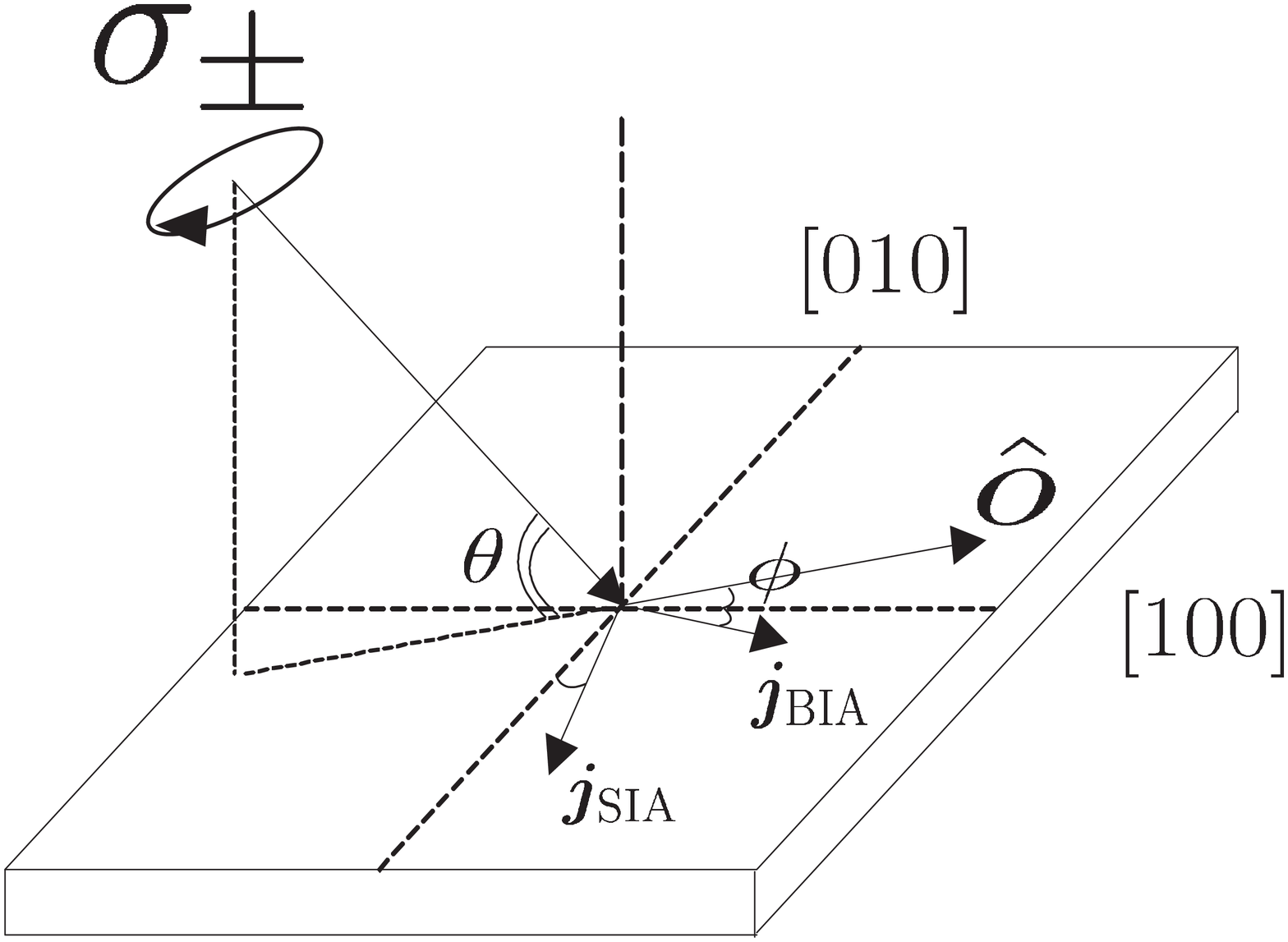}}
\caption{ \label{f0} SIA- and BIA-induced photocurrents appearing
under oblique incidence of circularly-polarized light.}
\end{figure}

Let us now consider optical excitation of a QW by
circularly-polarized light~(see Fig.~\ref{f0}).
Here we investigate the photocurrent arising due to asymmetry of
the carrier distribution in ${\bm k}$-space at the moment of
creation. It is different from another photocurrent caused by
carrier momentum re-distribution during the process of spin
relaxation. The latter, so-called spin-galvanic effect, was
considered in Refs.~\onlinecite{IL-GP_JETP}
and~\onlinecite{Ganichev_Nature}. These two photocurrents can be
separated in the time-resolved experiments: after switching-off
the light source, the former decays with the momentum relaxation
time while the latter disappears within the spin relaxation time.

The electric current is expressed in terms of the velocity
operators and spin density matrices for electrons and holes as
follows
\begin{equation}
\label{j} {\bm j} = e \sum_{\bm k} {\rm Tr} \left[ {\bm v}^{(e)}
({\bm k}) \: \rho^{(e)}({\bm k}) - {\bm v}^{(h)} ({\bm k}) \:
\rho^{(h)}({\bm k}) \right] \:,
\end{equation}
where $e$ is the electron charge.

The velocity matrix elements calculated on the wave
functions~(\ref{el_w_f}),~(\ref{h_w_f}) taking into account the
spin-orbit corrections~(\ref{H}), are given by
\begin{eqnarray}
\label{v} \left(v^{(e)}_i\right)_{mm'} &=& \left[ {\hbar k_i \over
m_e} + {m \over \hbar} (\beta_{xi} \cos{\Phi_{\bm k}} + \beta_{yi}
\sin{\Phi_{\bm k}}) \right] \delta_{mm'} \nonumber \\ &+& {i m
\over \hbar} (\beta_{yi} \cos{\Phi_{\bm k}} - \beta_{xi}
\sin{\Phi_{\bm k}}) (1 - \delta_{mm'})\:, \nonumber
\\
v^{(h)}_i &=& {k_i \over k} v_h(k), \:\:\:\:  v_h(k) = {1 \over
\hbar} {d E_h(k) \over dk}\:,
\end{eqnarray}
where $m_e$ is the electron effective mass.

Density-matrix equations taking into account both direct optical
transitions and elastic scattering give the following expressions
for linear in the light intensity values entering
into~Eq.(\ref{j})
\begin{eqnarray}
\label{rho} \rho^{(e,h)}_{nn'} = \pm {\pi \over \hbar} \tau_{e,h}
\sum_{\bar{n}} M_{n\bar{n}}M_{\bar{n}n'} &[\delta (E_n +
E_{\bar{n}} - \hbar \omega)& \\ + &\delta (E_{n'} + E_{\bar{n}} -
\hbar \omega)]& \nonumber \:.
\end{eqnarray}
Here $\tau_e$ and $\tau_h$ are the relaxation times of electrons
and holes (here we assume isotropic scattering), $\hbar \omega$ is
the photon energy, $M_{n\bar{n}}({\bm k})$ is the matrix element
of the direct optical transition between the subbands $n$ and
$\bar{n}$, and the energy dispersions $E_{e,h}({\bm k})$ are
reckoned inside the bands.

Calculations show that all odd harmonics of $\rho^{(e,h)} ({\bm
k})$ entering into~(\ref{j}) are proportional to the following
part of the sum
\begin{eqnarray}
\label{rho_odd} && \sum_{l=s,a} M_{ml}M_{lm'} \propto P_{circ}
\left( e p A_0 \over m_0 c \right)^2 \sin{\theta} \: (N Q_\pm)^2
\: m \\
&& \times \left\{ \delta_{mm'} [W_\pm \cos{(\Phi_{\bm k} - 2
\phi_{\bm k}
+ \phi)} \right. + W_\pm^2 \cos{(\Phi_{\bm k} - \phi)}] \nonumber \\
&& + i \: (1 - \delta_{mm'} )  \nonumber \\
&& \times \left. [W_\pm \sin{(\Phi_{\bm k} - 2 \phi_{\bm k} +
\phi)} + W_\pm^2 \sin{(\Phi_{\bm k} - \phi)}] \right\} \:.
\nonumber
\end{eqnarray}
Here $P_{circ}$ is the circular polarization degree, $\theta$ and
$\phi$ are the spherical angles of the light polarization
vector~(see Fig.~\ref{f0}), $p$ is the interband momentum matrix
element, $A_0$ is the light wave amplitude, $m_0$ is the free
electron mass, and
\begin{equation}
\label{Q} Q_+ = \int\limits_{-\infty}^{\infty} dz \: \varphi(z)
C(z)\:, \: Q_- = \zeta \int\limits_{-\infty}^{\infty} dz \:
\varphi(z) S(z)\:.
\end{equation}
The upper (lower) sign in Eq.~(\ref{rho_odd}) corresponds to
excitation into the odd (even) electron subbands.~\cite{zeta}

The characteristic spin splittings are usually very small,
therefore we consider a linear in $\beta$ regime. In this
approximation, SIA and BIA give independent contributions into the
photocurrent
\begin{equation}
\label{j_tot} {\bm j} (\omega) = {\bm j}_{\rm SIA} (\omega) + {\bm
j}_{\rm BIA} (\omega) \:,
\end{equation}
where ${\bm j}_{\rm SIA}$ and ${\bm j}_{\rm BIA}$ are linear in
$\beta_{\rm SIA}$ and $\beta_{\rm BIA}$, respectively. Assuming
the splitting $\Delta \to 0$ and calculating the reduced density
of states, we obtain from Eqs.~(\ref{j}) - (\ref{Q}) the
expressions for the interband circular photocurrents:
\begin{equation}
\label{j_fin} j_i (\omega) = - \beta_{li} \hat{o}_l \: P_{circ}
\left( e p A_0 \over m_0 \hbar c \right)^2 {e \over \hbar}  \:
G(k_\omega) \:.
\end{equation}
Here $i,l = x,y$, and $\hat{{\bm o}}$ is a projection of the unit
vector along the light propagation direction on the QW plane~(see
Fig.~\ref{f0}). The wave vector of the direct optical transition,
$k_\omega$, satisfies the energy conservation low
\begin{equation}
\label{k_omega}
E_e(k_\omega) + E_h(k_\omega) = \hbar \omega - E_g \:,
\end{equation}
where $E_e(k)$ is the parabolic electron dispersion without
spin-orbit corrections. We study the effects linear in $\beta$,
therefore Eq.~(\ref{j_fin}) is valid at $\hbar \omega - E_g^{QW}
\gg \beta k_\omega$, where $E_g^{QW} = E_g + E_{e1}(0) +
E_{h1}(0)$ is the fundamental energy gap corrected for the
energies of size-quantization. However for real systems, the
theory is valid even near the absorption edge.

The frequency dependence of the photocurrent is given by the
function $G(k)$
\begin{equation}
\label{G} G(k) = {k \over v(k)} {d \over dk} \left[ {F(k) u(k)
\over v(k)} \right] - {F(k) \over v(k)} \left[2 - {u(k) \over
v(k)} \right] \:,
\end{equation}
where
\begin{eqnarray}
\label{F&v&u} && F(k) = k [N(k) Q_\pm(k) W_\pm(k)]^2 \tau_e(k) \:, \\
&& v(k) = {\hbar k \over m_e} + v_h(k), \: u(k) = \left[ {\hbar k
\over m_e} + v_h(k) {\tau_h(k) \over \tau_e(k)} \right] \xi(k).
\nonumber
\end{eqnarray}

The first term in Eq.~(\ref{G}) appears because the direct
transitions to the upper (lower) spin branch take place at a wave
vector slightly smaller (larger) than $k_\omega$, and the second
term occurs because the two electron spin states with the same
$\bm k$ have different velocities.

The factor $\xi(k)$ depends on the form of a spin-orbit
interaction. It follows from Eqs.~(\ref{BIA}),~(\ref{SIA}) that,
for the BIA-induced spin-orbit interaction, $\Phi_{\bm k} = -
\phi_{\bm k}$, while, for SIA-dominance, $\Phi_{\bm k} = \phi_{\bm
k} - \pi/2$. Therefore one has
\begin{equation}
\label{xi} \xi_{\rm BIA}=1\:, \hspace{1cm} \xi_{\rm SIA}=1 -
1/W_\pm(k) \:.
\end{equation}
The difference appears because in the BIA-case the terms with
$W_\pm$ in Eq.~(\ref{rho_odd}) are the third harmonics of
$\phi_{\bm k}$ and, hence, do not contribute to the current. This
means that BIA and SIA create different current distributions of
optically-generated electrons.

This difference gives rise to non-equal frequency dependences of
the photocurrent. It is dramatic at the absorption edge, when
$\hbar \omega \geq E_g^{QW}$. For the ground hole subband $W_+(k)
\sim k^2$, and hence
\begin{equation}
\label{j_edge} j_{\rm BIA} \sim (\hbar \omega - E_g^{QW})^2 \:,
\hspace{1cm} j_{\rm SIA} \sim \hbar \omega - E_g^{QW} \:.
\end{equation}
This conclusion opens a possibility to distinguish experimentally
which kind of asymmetry, BIA or SIA, is dominant in the structure
under study. This could be done by studying the power, quadratic
or linear, in the dependence of the circular photocurrent on the
light frequency near the absorption edge. At higher photon
energies, the spectra are also different due to $k$-dependence of
the functions $W_\pm$ [see Eq.~(\ref{xi})].

\section{Results and Discussion}

Eqs.~(\ref{j_tot} - \ref{xi}) describe the contributions to the
circular photocurrent due to interband optical transitions. It is
seen that the symmetry of the system determines the direction of
the current. Indeed, according to Eq.~(\ref{j_fin}), $j_i (\omega)
\propto \beta_{li} \hat{o}_l$, i.e. (i)~the current appears only
under oblique light incidence and (ii)~for SIA the current  ${\bm
j}$ is perpendicular to $\hat{{\bm o}}$, while for BIA the angle
between ${\bm j}$ and $\hat{{\bm o}}$ is twice larger than the
angle between the axis [100] and $\hat{{\bm o}}$, see
Fig.~\ref{f0}.

\begin{figure*} \epsfxsize=3in \epsfysize=4in
\centering{\epsfbox{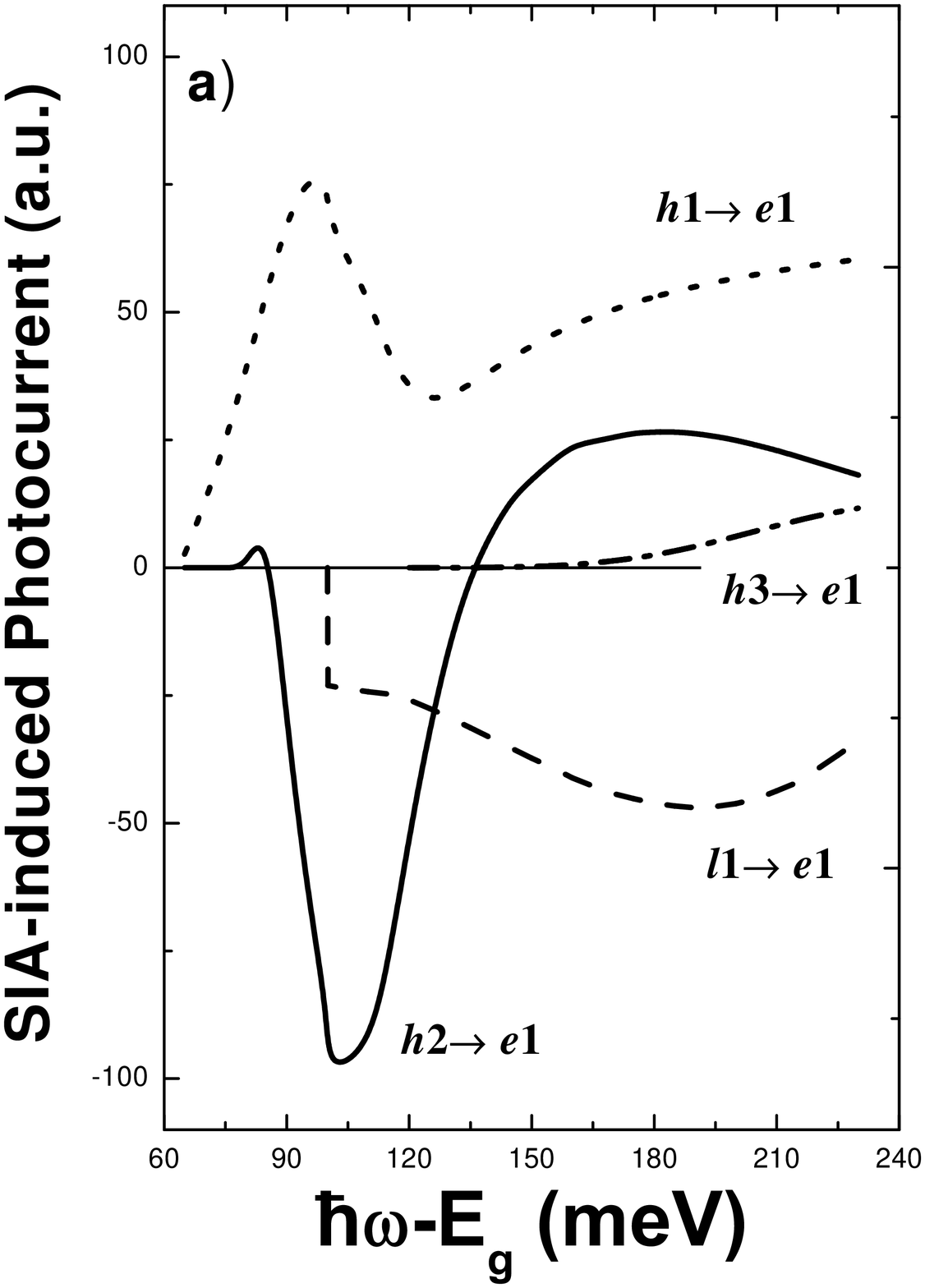}}  \epsfxsize=3in  \epsfxsize=3in
\epsfysize=4in \centering{\epsfbox{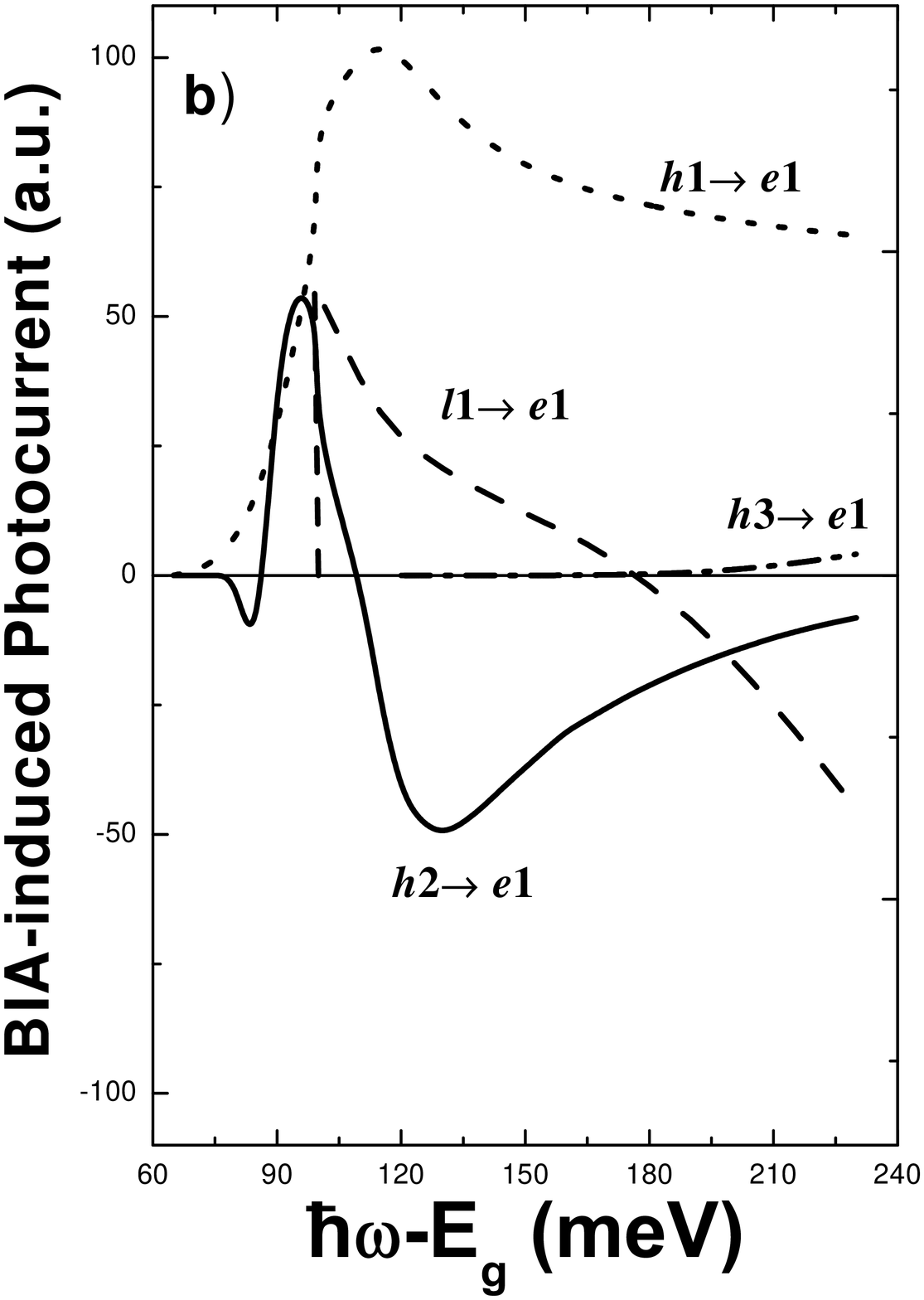}} \caption{\label{f2}
Partial contributions to the circular photocurrent for direct
interband transitions. Spin splitting of the electron states is
due to SIA (a) or BIA (b).}
\end{figure*}

\begin{figure}
\epsfxsize=3in \epsfysize=4in \centering{\epsfbox{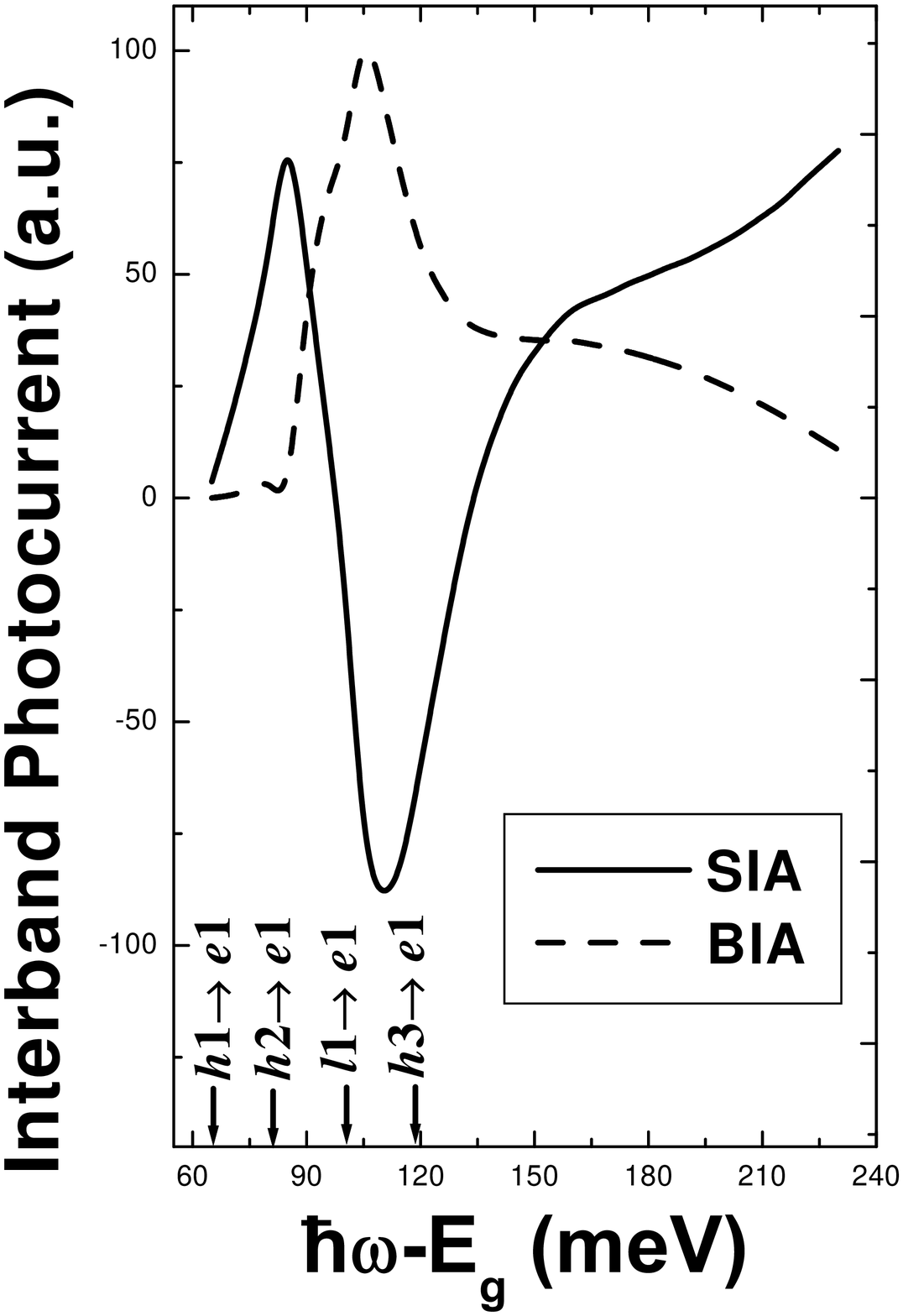}}
\caption{ \label{f1} Spectrum of the interband circular
photocurrent due to SIA (solid line) and BIA (dashed line) in a
100-\AA{} wide QW. The arrows indicate the absorption edges for
the four optical transitions.}
\end{figure}

The spectrum of the photocurrent is determined by the function
$G(k_\omega)$. Fig.~\ref{f2} presents the partial contributions to
this function for both  SIA and BIA calculated for a 100-\AA{}
wide QW with infinitely-high barriers. The interband transitions
between the $h1$, $h2$, $l1$ and $h3$ hole subbands and the ground
electron subband, $e1$, are taken into account. The effective
masses of the electron, heavy- and light-holes are chosen to
correspond to GaAs: $m_e=0.067 m_0$, $m_{hh}=0.51 m_0$,
$m_{lh}=0.082 m_0$. The momentum relaxation times are assumed to
be related by $\tau_h =2 \tau_e$ and independent of the carrier
energies.~\cite{tau}

The edge behavior of the photocurrent is due to the $h1 \to e1$
transitions. One can see the linear and quadratic raising of the
current near the absorption edge in accordance to
Eq.~(\ref{j_edge}). At higher energies, the spectra are determined
mainly by the $h1 \to e1$ and $h2 \to e1$ transitions. The
difference between BIA and SIA spectra is crucial: although in
both cases the current for the $h2 \to e1$ transitions is mainly
negative and has a minimum, in the BIA-induced photocurrent these
transitions give twice smaller contribution than $h1 \to e1$,
while for SIA-dominance they give the main contribution.

The total circular photocurrent caused by the four kinds of
optical transitions is presented in Fig.~\ref{f1}. Arrows in
Fig.~\ref{f1} indicate the points where the transitions start. One
can see that the BIA-induced circular photocurrent has a peak in
the spectrum, while in the SIA case the dip is present around the
same point. This photon energy corresponds to excitation of
carriers with $k \approx 2/a$, where $a$ is the quantum well
width. At this point the $h1$ and $h2$ energy dispersions have an
anti-crossing. This results in a transformation of the hole wave
functions and, hence, substantial changes in the dependence $G(k)$
Eq.~(\ref{G}).

The main feature of Fig.~\ref{f1} is that the BIA-photocurrent has
no sign change in the given energy domain, while in the SIA-case
it has a sign-variable spectrum. This makes possible to determine
the structure symmetry by means of the photogalvanic measurements.

\section{Concluding Remarks}
The situations are possible when the both types of asymmetry are
present. The absolute value of the current in the case $\beta_{\rm
SIA} \cdot \beta_{\rm BIA} \neq 0$ is given by
\begin{equation}
\label{j_tot_modul} j (\omega) = \sqrt{j_{\rm BIA}^2(\omega) +
j_{\rm SIA}^2(\omega) \mp 2 j_{\rm BIA}(\omega) j_{\rm
SIA}(\omega) \sin{2 \phi}} \:,
\end{equation}
where $\phi$ is the angle between $\hat{{\bm o}}$ and the axis
[100]. The upper (lower) sign in Eq.~(\ref{j_tot_modul})
corresponds to $\beta_{\rm BIA} \cdot \beta_{\rm SIA} > 0$ ($<
0$). The direction of the photocurrent is given by the angle
$\chi$ between $[1\bar{1}0]$ and ${\bm j}$:
\begin{equation}
\label{chi} \tan{\chi} = {j_{\rm SIA}(\omega) + j_{\rm
BIA}(\omega) \over j_{\rm SIA}(\omega) - j_{\rm
BIA}(\omega)}\tan{(\phi - \pi/4)}\:.
\end{equation}
The  angular dependence~(\ref{j_tot_modul}) occurs due to
simultaneous presence of Rashba and Dresselhaus fields
($\beta_{\rm SIA} \cdot \beta_{\rm BIA} \neq 0$) by analogy with
the $\phi_{\bm k}$-dependence of the spin splitting. It should be
noted that $j_{\rm SIA}$ and $j_{\rm BIA}$ have different
excitation spectra that causes complicated $\omega$-dependences of
the total photocurrent absolute value and direction,
Eqs.~(\ref{j_tot_modul},~\ref{chi}).

The analysis shows that, under anisotropic scattering, the
``interference'' terms $\beta_{\rm BIA} \cdot \beta_{\rm SIA} /
(\beta_{\rm BIA}^2 + \beta_{\rm SIA}^2)$ appear in the total
photocurrent~Eq.~(\ref{j_tot}) even in the linear in $\beta$'s
regime. This is caused by coupling of the Fourier harmonics of the
velocity operator and density matrix in~Eq.~(\ref{j}).

In conclusion, we have developed a theory of the interband
circular photogalvanic effect in QWs. The two cases when either
BIA or SIA dominates have been considered. It is shown that BIA
and SIA result in different photocurrents in QWs. The difference
is dramatic at the interband absorption edge, where the
photocurrent raises linearly or quadratically with increasing the
photon energy under the SIA or BIA dominance, respectively. At
higher $\hbar \omega$, the photocurrent excitation spectra also
have absolutely different shapes. This makes the interband
circular photogalvanic effect a unique high-sensitive tool for
investigation of symmetry and spin properties of QWs.

\section*{Acknowledgments}
Author thanks E.L.~Ivchenko for helpful discussions and
S.D.~Ganichev for critical reading of the manuscript. This work is
financially supported by the RFBR, DFG, INTAS, and by the
Programmes of Russian Ministry of Science and  Presidium of RAS.


\newpage
\begin{references}
\bibitem{spintronics} S. A. Wolf, D. D. Awschalom, R. A. Buhrman, J. M. Daughton, S. von Moln$\acute{a}$r, M. L. Roukes, A. Y. Chtchelkanova, and D. M. Treger,   Science {\bf 294}, 1488 (2001).

\bibitem{spin_transistor} S.~Datta and B.~Das, \apl {\bf 56}, 665 (1990).

\bibitem{Silva92} E. A. de Andrada e Silva,  \prb {\bf 46}, 1921 (1992).

\bibitem{SdHspin} D.~Stein, K.~von Klitzing, and G.~Weimann,  \prl {\bf 51}, 130 (1983).

\bibitem{DK} M.I.~D'yakonov and V.Yu.~Kachorovskii, Sov. Phys. Semicond. {\bf 20}, 110 (1986).

\bibitem{Jusserand92}B. Jusserand, D. Richards, H.~Peric, and B. Etienne,  \prl {\bf 69}, 848 (1992).

\bibitem{WLspin} P. D. Dresselhaus, C. M. A. Papavassiliou, R. G. Wheeler and R. N. Sacks,  \prl {\bf 68}, 106 (1992).




\bibitem{Review}N.\,S.~Averkiev, L.\,E.~Golub and M.~Willander,  J. Phys.: Condens. Matter {\bf 14}, R271 (2002).

\bibitem{ELGE} E.L. Ivchenko and G.E. Pikus, {\em Superlattices and Other Heterostructures. Symmetry and Optical Phenomena},
Springer Series in Solid State Sciences, Vol. 110, Springer-Verlag, Heidelberg, 1995; 2nd ed., 1997. Ch.~10.

\bibitem{Ganichev_last}S.D.~Ganichev, E.L.~Ivchenko, S.N.~Danilov, J.~Eroms, W.~Wegscheider, D.~Weiss, and W.~Prettl, Phys. Rev. Lett. {\bf 86}, 4358 (2001).

\bibitem{Ganichev_Nature}S.D.~Ganichev E.L.~Ivchenko, V.V.~Bel'kov, S.A.~Tarasenko, W.~Wegscheider, D.~Weiss, and W.~Prettl, Nature {\bf 417}, 153 (2002).

\bibitem{MPP}I.A.~Merkulov, V.I.~Perel', and M.E.~Portnoi, Sov. Phys. JETP {\bf 72}, 669 (1991).

\bibitem{zeta}Expressions for the real coefficients $N$, $W_\pm$, $\zeta$, $Q_\pm$, and for the functions $C(z)$ and $S(z)$ are given in Ref.~\onlinecite{MPP} in the approximation of infinitely-high barriers.

\bibitem{IL-GP_JETP}E.L.~Ivchenko, Yu.B.~Lyanda-Geller and G.E.~Pikus, Sov. Phys. JETP {\bf 71}, 550 (1990).


\bibitem{tau} The contribution of the only allowed transition $l1 \to e1$ is proportional to ($\tau_h - \tau_e$) at $k=0$, therefore we take different values for $\tau_e$ and $\tau_h$ in order to obtain a step-like behaviour of the spectrum. It is clearly seen in the corresponding curves in Figs.~\ref{f2}a and~\ref{f2}b.

\end{references}
\end{document}